\documentclass[9pt]{article}
\usepackage[preprint]{spconf}
\usepackage{amsmath,amstext}
\usepackage{amssymb}
\usepackage{graphicx}
\usepackage{cite,url}
\usepackage{caption}
\copyrightnotice{\copyright2019 IEEE}
\toappear{To appear at ICASSP 2019}

\title{ACOUSTIC IMPULSE RESPONSES FOR WEARABLE AUDIO DEVICES}

\name{Ryan M. Corey, Naoki Tsuda, and Andrew C. Singer\thanks{This material is based upon work supported by the National Science Foundation Graduate Research Fellowship Program under Grant Number DGE-1144245.}}
\address{University of Illinois at Urbana-Champaign}

\begin{document}

\maketitle

\begin{abstract}
\small
We present an open-access dataset of over 8000 acoustic impulse
from 160 microphones spread across the body and affixed to wearable accessories. The data can be used
to evaluate audio capture and array processing systems
using wearable devices such as hearing aids, headphones, eyeglasses,
jewelry, and clothing. We analyze the acoustic transfer functions
of different parts of the body, measure the effects of clothing worn
over microphones, compare measurements from a live human subject to
those from a mannequin, and simulate the noise-reduction
performance of several beamformers. The results suggest that arrays
of microphones spread across the body are more effective than those
confined to a single device.
\end{abstract}


\section{Introduction}

Thanks to advances in transducer technology, such as tiny digital MEMS microphones \cite{zwyssig2013speech},
multiple audio sensors can be embedded in wearable devices such as watches, headphones, eyeglasses, and other
accessories. These microphones could be combined to perform array processing such as beamforming, localization,
and source separation \cite{brandstein2013microphone,gannot2017consolidated,vincent2018audio}.
A wearable array with many microphones spread over a wide area would
offer greater spatial resolution than the small arrays embedded in
most hearing aids, headsets, and mobile phones today. Wearable
microphone arrays could dramatically improve performance in assistive listening \cite{Doclo2008,doclo2015magazine}, augmented reality \cite{valimaki2015assisted},
and machine perception applications.

There have been several wearable array designs reported
in the literature, including helmets \cite{scanlon2008helmet,gillett2009head,calamia2017helmet},
eyeglasses \cite{soede1993assessment,levin2016near}, and vests \cite{widrow2003microphone,stupakov2009cosine}.
However, these designs have been restricted to small areas of the
body and the literature offers little guidance about how microphone
placement affects performance. Furthermore, there is little publicly
available data, such as impulse response measurements, that can be used to design wearable arrays and test multimicrophone processing algorithms.

\begin{figure}
\begin{centering}
\includegraphics{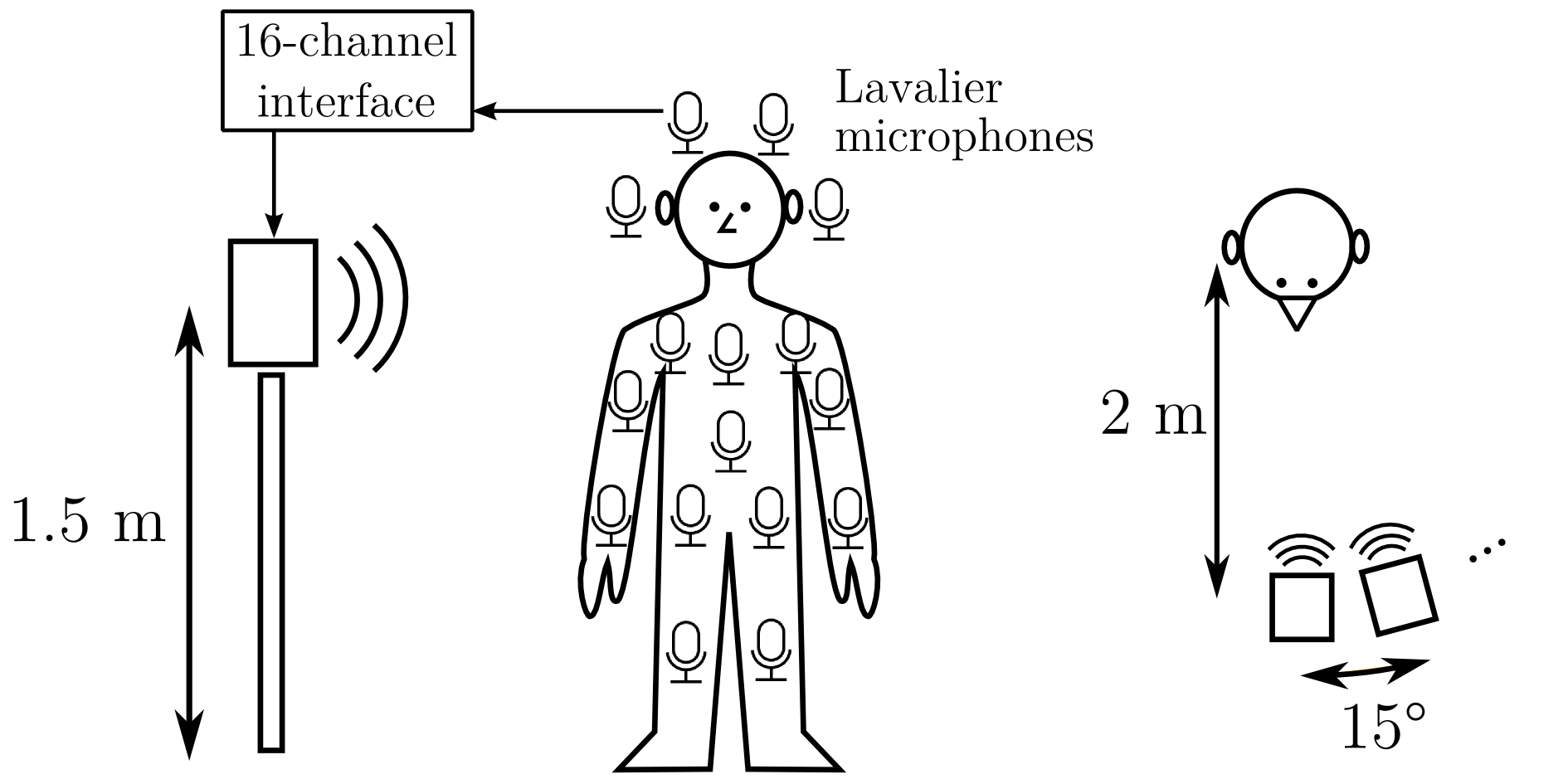}
\par\end{centering}
\caption{\footnotesize {\label{fig:setup}Impulse responses were measured using a studio monitor
and 16 microphones placed at 80 positions on the body and 80 positions
on wearable accessories. Test signals were captured from 24 angles.}}
\end{figure}

Multimicrophone impulse response datasets, such as \cite{wen06mardy,ono2013sisec,hadad2014multichannel},
are used to simulate sound propagation and evaluate reverberant source separation and beamforming algorithms.
There is abundant publicly available data on head-related transfer
functions (HRTF), which characterize directional filtering by the ears
\cite{blauert1997spatial}. HRTF datasets,
such as \cite{gardner1995hrtf,algazi2001cipic}, usually only include responses at the 
ear canals and sometimes at hearing-aid earpieces \cite{kayser2009database}. 
To simulate and evaluate wearable audio systems, researchers could use impulse responses measured with microphones placed all across the body. Note that whereas HRTFs are often used in human perceptual applications---for example, to create virtual sources
in a listener's auditory environment \cite{valimaki2015assisted}---these body-related transfer functions (BRTFs) are not directly related to human hearing. Rather, they help machines to localize,
separate, and enhance real-world sound sources, and could be used alongside HRTFs in listening enhancement applications. 

Here we present a new dataset \cite{brtf} of acoustic impulse responses measured at 160 sensor positions across the body and various wearable accessories. Version 1 of the wearable microphone dataset contains about 8000 measurements with one human subject, one mannequin, five head-mounted accessories and six types of outerwear. The data and documentation is available through the Illinois Data Bank\footnote{\url{https://doi.org/10.13012/B2IDB-1932389_V1}}, an open-access data archival service maintained by the University of Illinois at Urbana-Champaign. 

The wearable microphone dataset can be used to characterize the acoustic effects of the body on wearable audio devices and to simulate microphone arrays for applications such
as hearing aids, augmented reality, and human-computer
interaction.  In this paper, we analyze this data to describe the acoustic effects of different body parts, evaluate the mannequin as a human analogue, and compare the attenuation
of different clothing worn over microphones. Finally, we use the dataset
to assess designs of wearable microphone arrays for a beamforming
application. 

\section{Impulse Response Measurements}

The measurement setup is shown in Fig. \ref{fig:setup}. The impulse
responses were measured in an acoustically treated recording space
in the Illinois Augmented Listening Laboratory. Each half-second impulse response was computed from a ten-second linear sweep repeated three times from a studio
monitor, captured by 16 Countryman B3 omnidirectional
lavalier microphones, and digitized at 24 bits and 48 kHz by a Focusrite
Scarlett audio interface. After each sequence of sweeps, the subject
was rotated to capture impulse responses from a total of 24 source
angles. The microphones were then moved to new positions
and the measurements were repeated.


The human subject is 181 cm tall with a head circumference of 61 cm.
The hollow plastic mannequin, designed for displaying clothing, is
183 cm tall with a 56 cm head circumference. Since
the mannequin head has unnaturally small ears, a soft plastic replica
ear was affixed to each side of the head. These replica ears are not
intended to have realistic HRTFs, since HRTF data from realistic head
simulators and real humans is already readily available.

\begin{figure}
\begin{centering}
\hfill{}\includegraphics[height=4cm]{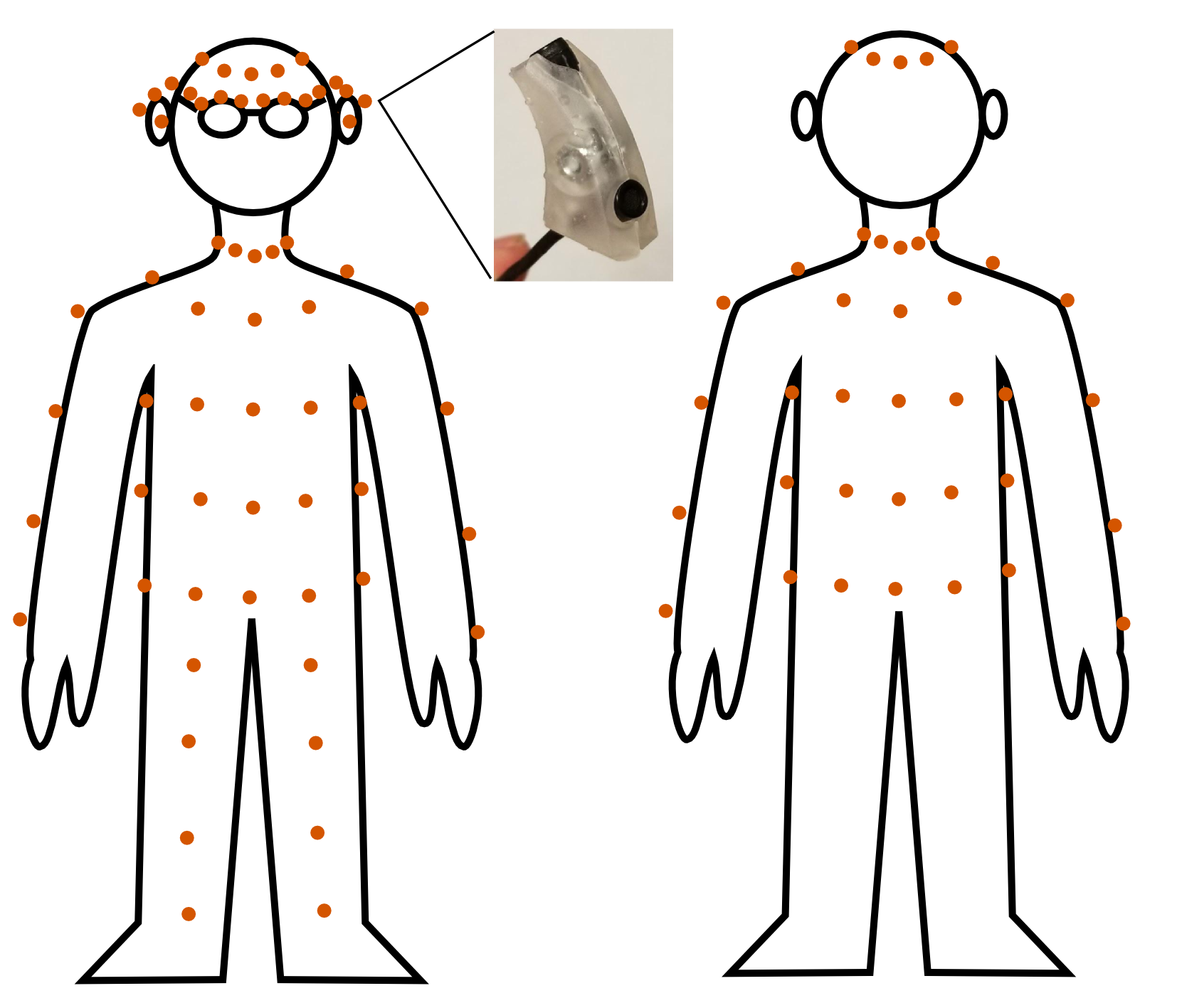}\hfill{}\includegraphics[height=4cm]{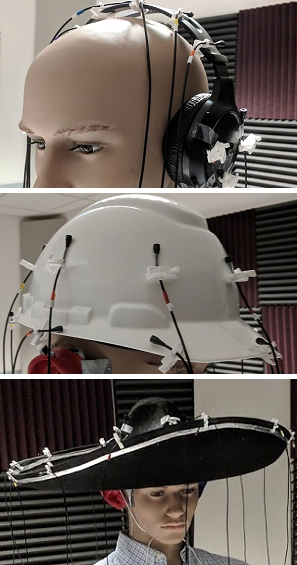}\hfill{}
\par\end{centering}
\caption{\label{fig:Microphone-placement}Left: Impulse responses
were measured at 80 positions on the body, including one microphone affixed to each ear and four in behind-the-ear shells.
Right: Wearable accessories with 16 microphones.}
\end{figure}

The BRTF data includes 80 microphone positions on the body, shown
in Fig. \ref{fig:Microphone-placement}. One microphone was placed
just outside of each ear canal and affixed using medical tape. These
microphones capture approximate HRTFs and can be used to simulate
binaural signal processing algorithms such as spatial-cue-preserving
beamformers \cite{doclo2006theoretical,marquardt2016development}.
Four microphones were mounted in a pair of custom-made
behind-the-ear (BTE) shells similar to those used in many hearing aids.
Ten were attached to a pair of eyeglasses and
the remaining 64 microphones were clipped onto the subject's clothing.

Since a wearable microphone array might be covered by clothing, the
torso measurements were repeated with different outerwear including
a t-shirt, cotton dress shirt, heavy cotton sweatshirt, polyester
pullover, wool coat, and leather jacket.

These BRTF measurements are supplemented by impulse responses from
wearable accessories. Since many previously reported wearable arrays
are mounted on the head, measurements were collected using over-the-ear
headphones, a baseball cap, a hard hat, a hat with a 40 cm flat brim,
and a hat with a 60 cm curved brim, each with 16 microphones.

\section{Acoustic Transfer Functions}

\subsection{Effects of the body}

\begin{figure}
\begin{centering}
\includegraphics{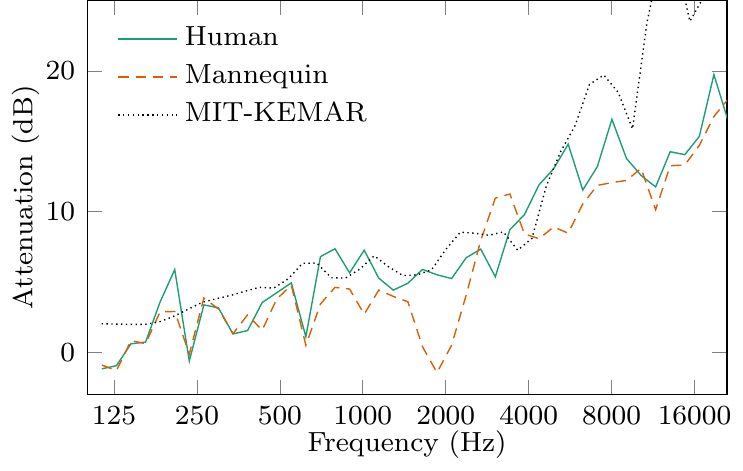}
\par\end{centering}
\caption{\label{fig:headshadow}Interaural level differences for sources to the left and right of the subject. The dotted curve is from the
MIT KEMAR dataset \cite{gardner1995hrtf}.}
\end{figure}

\begin{figure}
\begin{centering}
\includegraphics{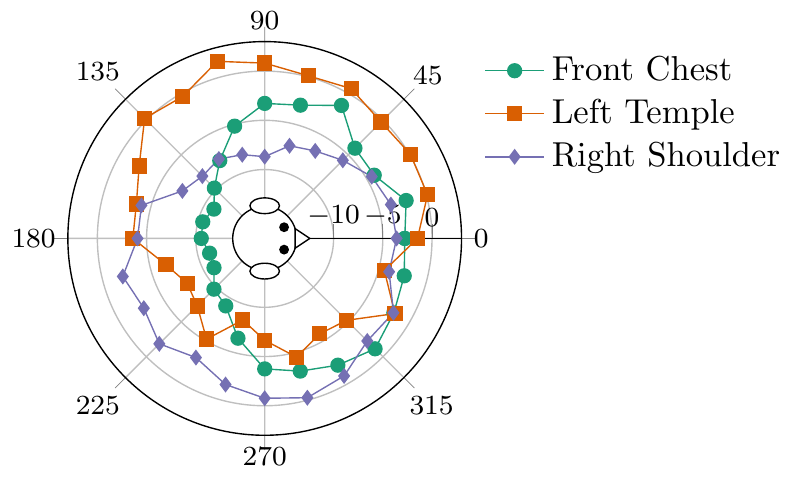}
\par\end{centering}
\caption{\label{fig:Directivity-of-microphones}Overall power, in dB relative
to a free-space microphone, received by three microphones on the human
subject.}
\end{figure}

The acoustic effects of the head, which humans use to localize sound, have been well studied \cite{blauert1997spatial}.
A microphone in the left ear will capture more energy
from sources on the left than sources on the right, especially at
high frequencies. This interaural level difference is shown in Fig.
\ref{fig:headshadow}. The human head has a slightly stronger acoustic
shadow effect than the plastic mannequin head. The head-shadow effect
measured in the treated recording space is slightly weaker than fully-anechoic
KEMAR data from \cite{gardner1995hrtf}.

The rest of the body has similar shadowing effects, which causes omnidirectional wearable 
microphones to have directional responses, as shown in Fig. \ref{fig:Directivity-of-microphones}. 
A microphone on the front of the chest receives about
8 dB less sound energy from sources behind the wearer. Microphones
on the temple and shoulder are shadowed from the side but not from
the front.

\begin{figure}
\begin{centering}
\includegraphics{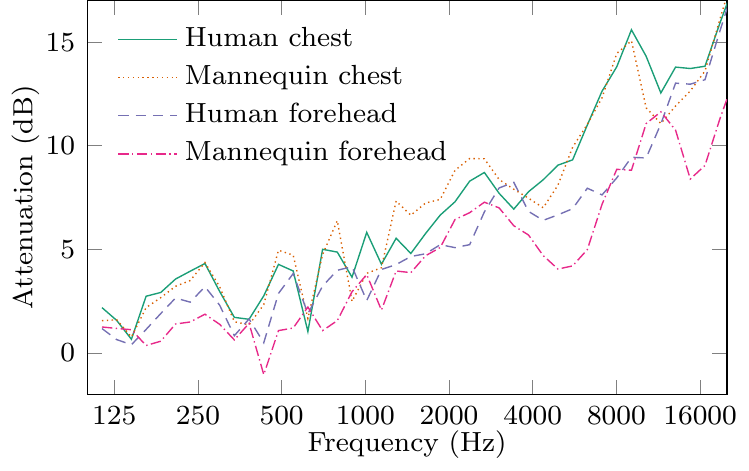}
\par\end{centering}
\caption{\label{fig:Attenuation-by-the}Average attenuation by the body for sources on the opposite side of the body from each microphone.}
\end{figure}

The body-related shadow effect varies with frequency and body part.
For both the human and mannequin, the shadow effect was strongest
for the the upper chest and weakest for the forehead, although the
differences between body parts are small compared to variations across
frequency. Fig. \ref{fig:Attenuation-by-the} shows the average difference
in transfer function magnitude between the sources nearest to and
farthest from each microphone on the upper chest and forehead. The
transfer functions for the human and mannequin are similar in magnitude,
suggesting that inexpensive plastic mannequins can be used as human
analogues in wearable-microphone experiments.

\subsection{Effects of clothing}

\begin{figure}
\begin{centering}
\includegraphics{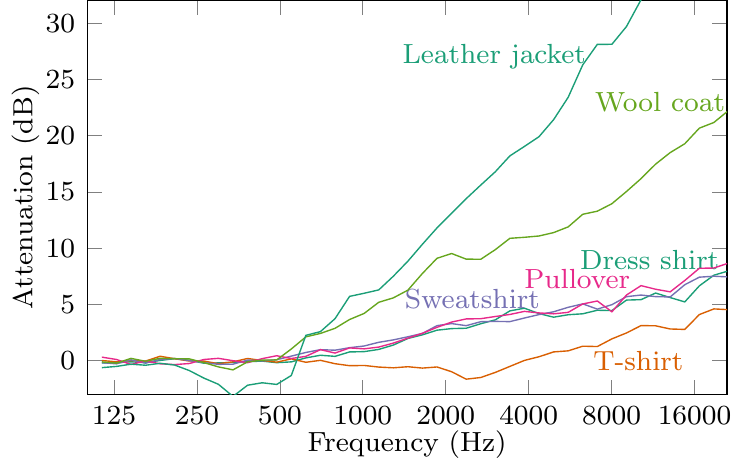}
\par\end{centering}
\caption{\label{fig:clothing}Average attenuation due to clothing for the 16
microphones on the mannequin torso.}
\end{figure}

In many wearable-audio applications, microphones might be worn in,
on, or under clothing. In the HRTF literature, it has been shown that
hair, eyeglasses, and hats have small but measurable effects on acoustic
transfer functions to the ear \cite{wersenyi2005differences,riederer2005hrtf,treeby2007effect}
but do not significantly affect human localization performance \cite{riederer2005hrtf,wersenyi2017comparison}.
The strongest effects are from curly hairstyles that cover the pinna
and wide-brimmed hats that reflect sounds from below into the ear
and sounds from above away from the ear \cite{riederer2005hrtf}.
Clothing worn on the torso has little effect on HRTFs---at
most, it changes the strength of multipath reflections from sources
below the listener \cite{riederer2005hrtf}---but would of
course have a strong effect on BRTFs. 

The attenuation due to different clothing, averaged over all microphones
on the torso, is shown in Fig. \ref{fig:clothing}. All garments attenuate higher frequencies, but the degree of attenuation depends
on the type of clothing. The t-shirt has the smallest effect, up to
5 dB at 20 kHz. The light cotton dress shirt, heavy cotton sweatshirt,
and polyester pullover have nearly identical attenuation effects.
The wool coat and leather jacket have strong high-frequency attenuation, suggesting
that wearable audio devices might be less useful when covered by heavy outerwear. Note that the leather jacket appears to slightly amplify sound around 200--600 Hz in this recording setup; the effect was consistent across all microphones.

\section{Application to Beamforming}

Microphone arrays are often used for beamforming, that is, to isolate a desired source and remove
unwanted noise \cite{van1988beamforming,Doclo2008,gannot2017consolidated}.
A wearable array with many microphones spread across the body could perform 
stronger noise reduction than the small arrays included in many audio devices today.
The wearable microphone dataset developed here can be used to study how performance scales with array size in a wearable application and how such arrays should be designed.

\subsection{MVDR beamformer}

Let $s[n]\in\mathbb{R}$ be a sequence of speech samples emitted from
a nonmoving source of interest. Let $\mathbf{a}[n]\in\mathbb{R}^{M}$
be an $M$-dimensional impulse response 
from the source to each of $M$ microphones in an array.
Let $\mathbf{z}[n]\in\mathbb{R}^{M}$ be an unwanted noise sequence.
Assuming linear time-invariant propagation, the sampled recorded signal is
\begin{equation}
\mathbf{x}[n]=\sum_{k=-\infty}^{\infty}\mathbf{a}[k]s[n-k]+\mathbf{z}[n].\label{eq:convolution}
\end{equation}
In the frequency domain, (\ref{eq:convolution}) can be written 
\begin{equation}
\mathbf{X}(\omega)=\mathbf{A}(\omega)S(\omega)+\mathbf{Z}(\omega),
\end{equation}
where $\mathbf{A}(\omega)$ is the discrete-time acoustic transfer
function vector and $\mathbf{X}(\omega)$, $S(\omega)$, and $\mathbf{Z}(\omega)$
are discrete-time Fourier transforms of the corresponding sequences.

If $\mathbf{z}[n]$ is a wide-sense stationary random process with
power spectral density $\mathbf{R}(\omega)$, then the output
$y[n]$ of a minimum-variance distortionless-response (MVDR) beamformer
is given in the frequency domain by 
\begin{equation}
Y(\omega)=A_{1}(\omega)\frac{\mathbf{A}^{H}(\omega)\mathbf{R}^{-1}(\omega)}{\mathbf{A}^{H}(\omega)\mathbf{R}^{-1}(\omega)\mathbf{A}(\omega)}\mathbf{X}(\omega).
\end{equation}

This beamformer minimizes noise power subject to the constraint that
the output due to the target source has unity gain with respect to
microphone 1, which is near the left ear. In a binaural system, there would be a second
output with unity gain with respect to the right-ear microphone. This constraint
ensures that the target source sounds natural to the listener, although
any residual noise will be spatially and spectrally distorted \cite{doclo2006theoretical}.

The performance metric used in these experiments is the improvement
in signal-to-noise ratio (SNR) between input and output:
\begin{equation}
\Delta\text{SNR}=10\log_{10}\frac{\sum_{n}\left(d[n]-x_{1}[n]\right)^{2}}{\sum_{n}\left(d[n]-y[n]\right)^{2}},
\end{equation}
where $d[n]$ is the noise-free desired sequence. 

\subsection{Beamforming simulation}

\begin{figure}
\begin{centering}
\includegraphics{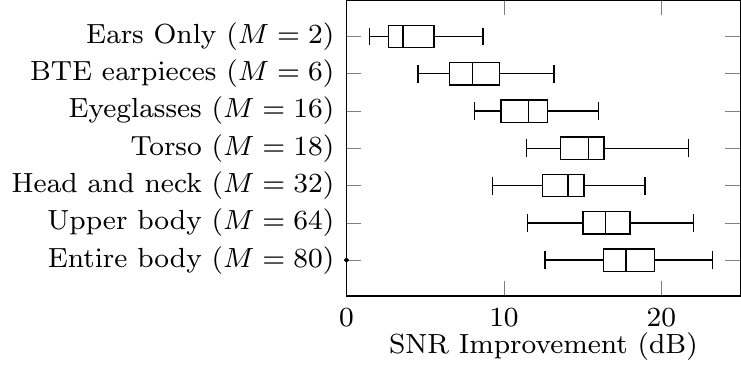}
\par\end{centering}
\caption{\label{fig:mic_number}Experimental results for MVDR beamforming on
the human subject with wearable arrays having different numbers of
microphones. All arrays include the reference microphones near the
ear canals. The box-and-whiskers plot indicates the quartiles of the simulated SNR improvements.}
\end{figure}

An MVDR beamformer was simulated using several wearable array configurations
with different numbers of microphones. For each of 100 trials, a target
source and five interference sources were randomly placed at six of
the 24 possible source locations. The source data was also randomly
chosen from a set of ten-second anechoic speech clips from the VCTK
corpus \cite{Veaux2017}. Since the source impulse responses are known,
an MVDR beamformer with more than six inputs could achieve near-perfect
performance by placing a null over each source. To prevent this overfitting,
the beamformer was designed using 32 ms windowed impulse responses
and diagonal loading about 10 dB below the average speech power.

The results of the beamforming experiment for different numbers of 
microphones are shown in Fig. \ref{fig:mic_number}.
Performance improves rapidly with the first few sensors as each new
input allows the beamformer to cancel an additional source. Larger
arrays offer more marginal improvements, helping to reduce residual
noise and compensate for transfer-function mismatch. The locations
of the microphones also affect performance: notice that the 18 microphones
on the ear canals and torso outperform 32 microphones on the head.
The microphones on the head are closely spaced, while those on the
torso are widely separated and also more strongly shadowed by the body.

\begin{figure}
\begin{centering}
\includegraphics{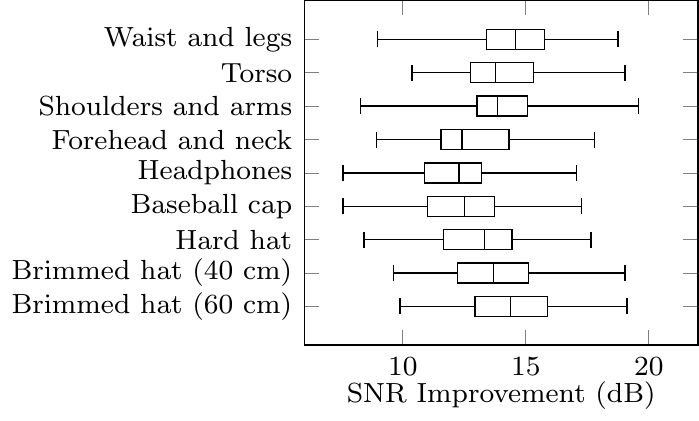}
\par\end{centering}
\caption{\label{fig:hats}Experimental results for MVDR beamforming on the
mannequin with different microphone configurations. Each array has
$M=18$ microphones, including the left and right reference microphones.}
\end{figure}

Fig. \ref{fig:hats} shows the performance of several arrays with
$M=18$ microphones, two of which are the left and right-ear reference microphones.
Comparing different head-mounted accessories, the largest hat provides
the best beamforming gain because of its spatial diversity. The microphones
attached to the over-the-ear headphones are too closely spaced to
provide much benefit at low frequencies and do not experience a strong
shadowing effect at high frequencies. The 60 cm hat is about as
effective as the lower-body array, which covers the largest 
area among the clothing-based arrays.

\section{Conclusions}

Many audio products, especially wearable devices such as hearing aids
and headsets, use relatively few microphones that are closely spaced.
The beamforming simulation suggests that performance could
be improved by using many microphones spread across
the body. For example, an array of 18 microphones across the torso
reduced noise by an average of about 2 dB more than an array of 18
microphones spaced across headphones. It also outperformed an array
of nearly twice as many microphones covering the head alone! The experiments
with clothing suggest that wearable microphones remain useful even
when covered by heavy shirts and sweaters, though wind-blocking coats
and jackets cause significant attenuation.

Further work is required to understand how acoustic transfer functions
vary between individuals. The wearable microphone dataset could be expanded in the future to include more human subjects and wearable devices. This data will allow researchers to simulate and compare different wearable array designs
and to develop new signal processing methods that take advantage of
larger arrays than are typically used today. 

\bibliographystyle{ieeetran}
\small \bibliography{../master_references}

\end{document}